\documentclass[aps,prl,reprint,groupedaddress]{revtex4-1}

\usepackage{amsmath,graphicx,color}
\usepackage{bm}

\newcommand{{\todo}}[1]{{\color{red} \bf #1}}
\newcommand{\pd}{{\phantom{\dagger}}}
\newcommand{\up}{\uparrow}
\newcommand{\dw}{\downarrow}

\begin{document}

\title{Topological Superconductivity in Skyrmion Lattices}

\author{Eric Mascot$^{1}$, Jasmin Bedow$^{1,2}$, Martin Graham$^{1}$, Stephan Rachel,$^{3}$, and Dirk K. Morr$^{1}$}
\affiliation{$^{1}$ Department of Physics, University of Illinois at Chicago, Chicago, IL 60607, USA}
\affiliation{$^{2}$ Lehrstuhl f\"{u}r Theoretische Physik I, Technische Universit\"{a}t Dortmund, 44221 Dortmund, Germany}
\affiliation{$^{3}$ School of Physics, University of Melbourne, Parkville, VIC 3010, Australia}

\maketitle
\nopagebreak

\textbf{Atomic manipulation and interface engineering techniques have provided a novel approach to custom-designing topological superconductors and the ensuing Majorana zero modes, representing a new paradigm for the realization of topological quantum computing and topology-based devices. Magnet-superconductor hybrid (MSH) systems have proven to be experimentally suitable to engineer topological superconductivity through the control of both the complex structure of its magnetic layer and the interface properties of the superconducting surface. Here, we demonstrate that two-dimensional MSH systems containing a magnetic skyrmion lattice provide an unprecedented ability to control the emergence of topological phases. By changing the skyrmion radius, which can be achieved experimentally through an external magnetic field, one can tune between different topological superconducting phases, allowing one to explore their unique properties and the transitions between them. In these MSH systems, Josephson scanning tunneling spectroscopy spatially visualizes one of the most crucial aspects underlying the emergence of topological superconductivity, the spatial structure of the induced spin-triplet correlations.}

The ability to create, control and manipulate topological superconducting phases is quintessential for the realization of topological quantum computing using the non-Abelian braiding statistics of Majorana zero modes \cite{nayak-08rmp1083}. These modes have been observed in one- \cite{mourik-12s1003,nadj-perge-14s602,ruby-15prl197204,pawlak-15npjqi16035,kim-18sa5251} and two-dimensional (2D) topological superconductors \cite{Wang18,Mac19,Men19}, with the latter also exhibiting chiral Majorana edge modes \cite{menard16-nc2040,Pal19,Wang20}. Magnet-superconductor hybrid (MSH) systems consisting of chains, islands or layers of magnetic adatoms deposited on the surface of conventional $s$-wave superconductors, have proven to be suitable experimental systems for (a) the creation of topological superconductivity using atomic manipulation \cite{kim-18sa5251} or interface engineering techniques \cite{Pal19}, and (b) the study of Majorana modes using scanning tunneling spectroscopy (STS). In particular, 2D MSH systems, with their topological invariant given by the Chern number, are predicted to exhibit a rich topological phase diagram \cite{Ron15,Li16,Rac17}. However, the experimental ability to tune between different topological phases in 2D MSH systems, essential for exploring the nature of topological superconductivity, has not yet been realized.

In the following, we demonstrate that the ability to tune between topological phases can be achieved in 2D MSH systems containing a magnetic skyrmion lattice by varying the skyrmion radius. As the latter can be experimentally controlled through the application of an external magnetic field \cite{Rom15}, the skyrmion MSH system presents an unprecedented opportunity to explore a rich phase diagram of topological superconducting phases, and the transitions between them. The underlying origin for the ability to control the topological phases lies in a spatially inhomogeneous Rashba spin-orbit (RSO) interaction that is induced by the magnetic skyrmion lattice. The induced RSO interaction images the local topological skyrmion charge -- the skyrmion number density -- and possesses a characteristic spatial signature in the zero-energy local density of states (LDOS) which can be observed at a topological phase transition as well as in the LDOS of chiral Majorana edge modes. Finally, we demonstrate that Josephson scanning tunneling spectroscopy can be employed to visualize one of the most fundamental aspects underlying the emergence of topological superconductivity, the existence of induced spin-triplet superconducting correlations. As 2D skyrmion MSH systems can be built with currently available experimental techniques, our results open new venues for the exploration and manipulation of topological superconductivity and Majorana zero modes.\\[-0.2cm]

\begin{figure}[t]
\centering
\includegraphics[width=8.5cm]{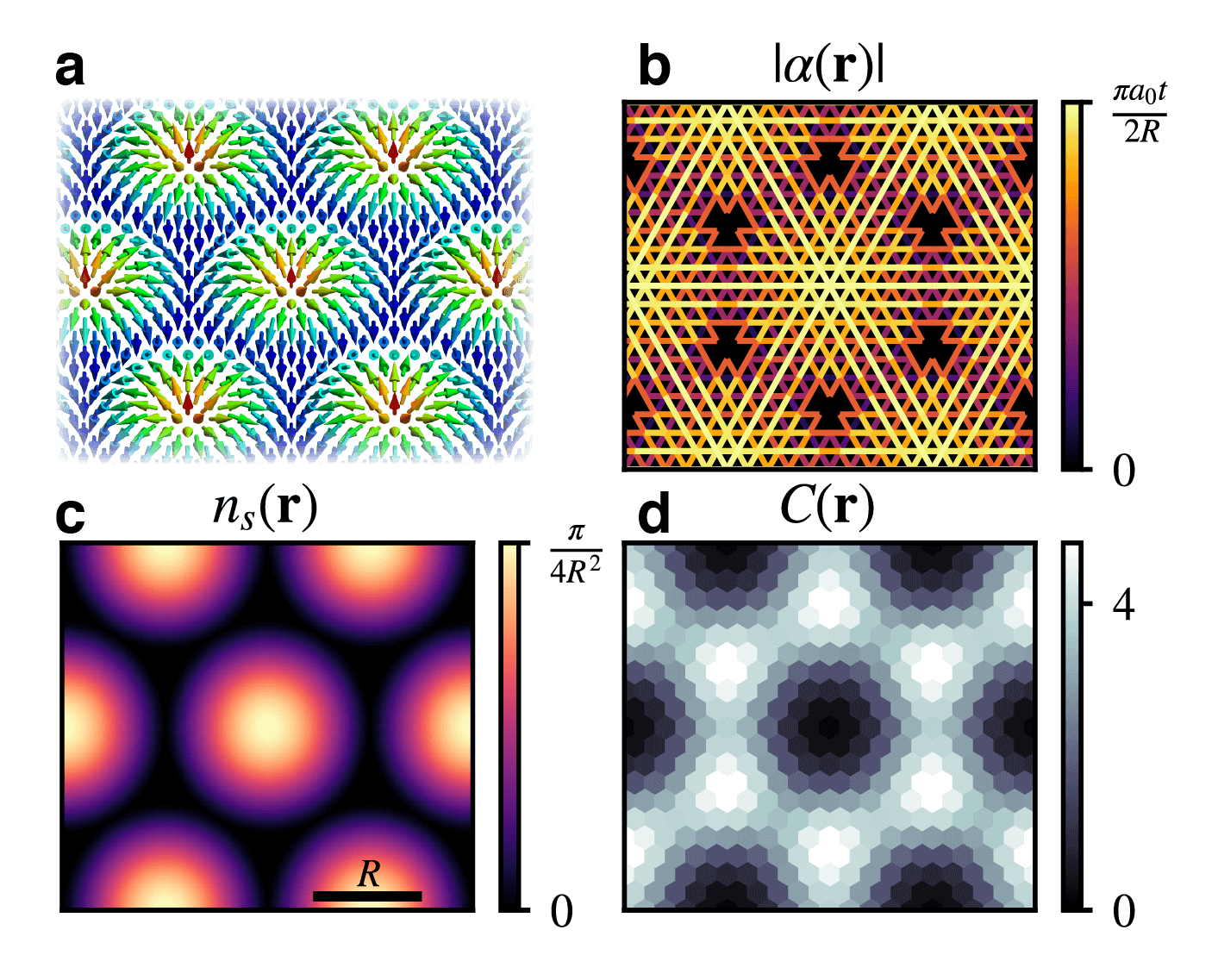}
\caption
{MSH system with a magnetic skyrmion lattice. {\bf a} Schematic picture of a skyrmion lattice. Spatial plot of {\bf b} the magnitude of the induced Rashba spin orbit interaction, $|\alpha({\bf r})|$, {\bf c} the skyrmion number density, $n_s({\bf r})$, and {\bf d} the Chern number density $C({\bf r})$ for skyrmion radius $R=5a_0$, and parameters $(\mu,\Delta,JS) = (-5,0.4,0.5)t$.}
\label{fig:fig1}
\end{figure}

\noindent {\large {\bf Results}}\\
\noindent {\bf Theoretical Model.}
We investigate the emergence of topological superconductivity in a 2D MSH system, in which a
magnetic skyrmion lattice (see Fig.~\ref{fig:fig1}{\bf a}) is placed on the surface of a conventional $s$-wave superconductor, as described by the Hamiltonian
\begin{align}
H &= \sum_{ {\bf r},{\bf r^\prime}, \sigma} \left( -t_{\bf rr^\prime} - \mu \delta_{\bf r,r^\prime} \right) c_{{\bf r} \sigma}^\dag c^\pd_{{\bf r^\prime} \sigma} + \Delta \sum_{\bf r} \left( c_{{\bf r} \up}^\dag c^\dag_{{\bf r} \dw} + {\rm H.c.} \right) \nonumber \\
& + J  \sum_{{\bf r},\alpha,\beta} {\bf S}_{\bf r} \cdot c_{{\bf r} \alpha}^\dag {\boldsymbol \sigma}_{\alpha \beta} c^\pd_{{\bf r} \beta} \ ,
\label{eq:H}
\end{align}
where $c_{{\bf r}\alpha}^\dag$ creates an electron at lattice site ${\bf r}$ with spin $\alpha$, and ${\bm \sigma}$ is the vector of spin Pauli matrices. We consider a triangular lattice with lattice constant $a_0$, chemical potential $\mu$, and hopping amplitude $-t_{\bf rr^\prime}=-t$ between nearest-neighbor sites only. $\Delta$ is the superconducting $s$-wave order parameter. The spatial spin structure of the skyrmion lattice is encoded in ${\bf S}_{\bf r}$ [see Supplementary Information (SI) section 1], which represents the direction of an adatom's spin located at site ${\bf r}$, and $J$ is its exchange coupling with the conduction electron spin. Note that the creation of Majorana modes in single skyrmions has previously been discussed in Refs.\cite{Yang16,Gar19}. As Kondo screening is suppressed by the full superconducting gap, the spins ${\bf S}_{\bf r}$ are taken to be classical vectors of length $S$. We assume that the triangular lattice of skyrmions is commensurate with the underlying triangular surface lattice, thus allowing the skyrmion radius $R$ to take integer and half-integer values of $a_0$. Note that in contrast to earlier studies of 2D MSH systems \cite{Ron15,Li16,Rac17}, the above Hamiltonian does not contain an intrinsic Rashba spin-orbit (RSO) interaction. Moreover, due to the broken time-reversal symmetry arising from the presence of magnetic moments, and the particle-hole symmetry of the superconducting state, the topological superconductor belongs to class D \cite{Ryu10}.

To characterize the topological superconducting phases of the system, we compute the topological invariant -- the Chern number $C$ -- given by
\begin{align}\label{eq:C}
 C & =  \frac{1}{2\pi i} \int_{\text{BZ}} d^2k \mathrm{Tr} ( P_{\bf{k}} [ \partial_{k_x} P_{\bf{k}}, \partial_{k_y} P_{\bf{k}} ] )  \nonumber \\
 P_{\bf{k}} & = \sum_{E_n(\bf{k}) < 0} |\Psi_n({\bf{k}}) \rangle \langle \Psi_n({\bf{k}})|
\end{align}
where $E_n({\bf{k}})$ and $|\Psi_n({\bf{k}}) \rangle$ are the eigenenergies and the eigenvectors of the Hamiltonian in Eq.(\ref{eq:H}), with $n$ being a band index, and the trace is taken over Nambu and spin-space. Further insight into the origin underlying the emergence of topological superconductivity in skyrmion MSH systems can be gained by considering the spatial structure of the skyrmion and Chern number densities, $n_s({\bf r})$ and $C({\bf r})$, respectively. The former is given by
\begin{align}\label{eq:skyrdensity}
 n_s({\bf r}) & =  \frac{1}{4\pi}  {\bf S(r)} \cdot \left[ \partial_x {\bf S(r)} \times \partial_y {\bf S(r)}  \right]
\end{align}
yielding a skyrmion number $n_s = \sum_{\bf r} n_s({\bf r})$.  The latter, $C({\bf r})$ \cite{Mas19,bianco-11prb241106} (see SI section 2), represents the real-space analog of the Berry curvature, and allows a real space calculation of the Chern number \cite{prodan11jpa113001,prodan17} $C = 1/N^2 \sum_{\textbf{r}}  C(\textbf{r})$ that coincides with that obtained from Eq.(\ref{eq:C}). \\[-0.2cm]

\begin{figure}[t]
\centering
\includegraphics[width=8.5cm]{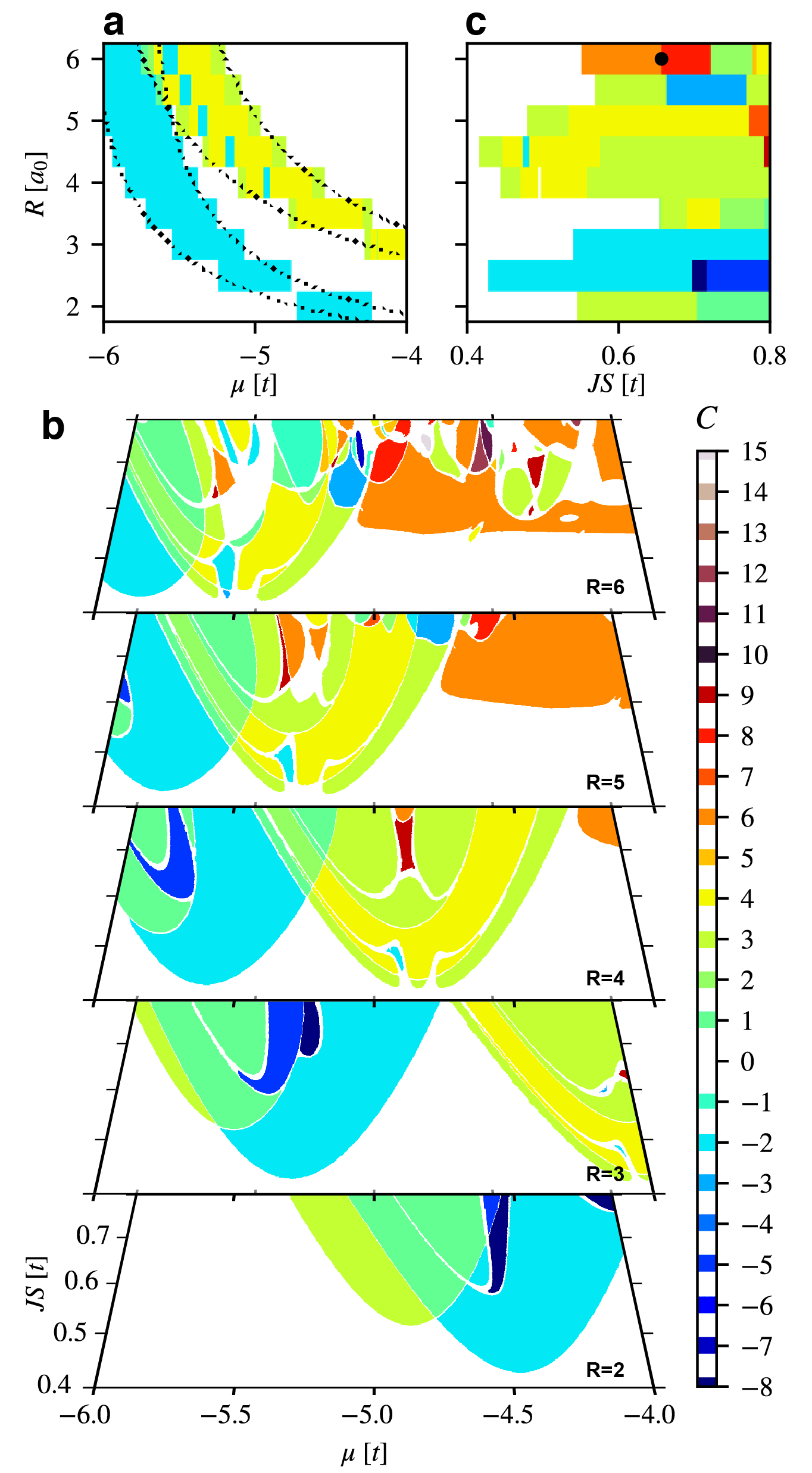}
\caption
{Topological phase diagrams of a skyrmion MSH system. Topological phase diagrams representing the Chern number, $C$, in the {\bf a} $(\mu,R)$ plane for $JS=0.5$t, {\bf b} $(\mu, JS)$ planes for various skyrmion radii $R$, and {\bf c} the $(JS,R)$ plane for $\mu=-5$t and $\Delta=0.4$t. Dashed lines in {\bf a} represent phase transition lines described by $\mu=A_i+B_i/R^2$ with constants $A_i,B_i$.}
\label{fig:fig2}
\end{figure}

\noindent {\bf Topological phase diagram.} A crucial aspect for the emergence of topological superconductivity in 2D skyrmion MSH systems is that the magnetic skyrmion lattice induces an effective, spatially varying Rashba spin-orbit interaction. To demonstrate this, we apply a unitary transformation \cite{Chen15} to the Hamiltonian in Eq.(\ref{eq:H}) (see SI section 2) that rotates the local spin ${\bf S}_{\bf r}$ to the ${\hat z}$ axis, yielding an out-of-plane ferromagnetic order and a spatially inhomogeneous RSO interaction, $\alpha({\bf r})$ (see Fig.\ref{fig:fig1}{\bf b}). $\alpha({\bf r})$ possesses the same spatial structure as the skyrmion number density, $n_s({\bf r})$, (see Fig.\ref{fig:fig1}{\bf c}) --  reflecting its origin in the local topological charge of the skyrmon lattice -- with its largest value, $\alpha_{max} = \pi a_0 t/(2R)$, in the center of the skyrmion, and a vanishing $\alpha({\bf r})$ at the corners of the skyrmion lattice Wigner-Seitz unit cell. The existence of a non-zero $\alpha({\bf r})$, of an out-of-plane ferromagnetic order in the rotated basis, and of a hard $s$-wave gap, satisfies all necessary requirements for the emergence of topological superconductivity \cite{Ron15,Li16,Rac17}, resulting in the rich topological phase diagram shown in Fig.~\ref{fig:fig2}.

The phase diagram in the $(\mu,R)$ plane (see Fig.~\ref{fig:fig2}{\bf a}) reveals the intriguing result that it is possible to tune a skyrmion MSH system between different topological phases by changing the skyrmion radius $R$, which can be experimentally achieved through the application of an external magnetic field \cite{Rom15}. This unprecedented ability arises from the facts that (a) varying the skyrmion radius leads to changes in the induced $\alpha(\bf r)$, and (b) in contrast to MSH systems with a homogeneous ferromagnetic structure, topological phase transitions in magnetically inhomogeneous MSH systems (as given here) are controlled not only by $\mu$ and $J$, but also by $\alpha$. Indeed, the results in Fig.~\ref{fig:fig2}{\bf a} reveal that the phase transition lines in the $(\mu,R)$ plane are determined by $\mu=A_i+B_i/R^2$ (see dashed lines) with constants $A_i,B_i$. Since $\alpha_{max} \sim 1/R$, our result suggests that the induced RSO interaction leads to an effective renormalization of the chemical potential \cite{Yang16}, thus facilitating the ability to tune between topological phases. This dependence of the phase transition lines on $R$ is also revealed when considering the phase diagrams in the $(\mu,JS)$ plane for different skyrmion radii (see Fig.~\ref{fig:fig2}{\bf b}). These phase diagrams show a very similar structure of topological phases for different $R$, with the phases moving to lower values of $\mu$ with increasing $R$. We note that the topological phases that are accessible through tuning of $R$ strongly depend on $JS$ (see Fig.~\ref{fig:fig2}{\bf c}): for sufficiently large $JS$, every change in the skyrmion radius by a half-integer leads to a change in the system's Chern number. Thus, a rich topological phase diagram can be accessed and explored through changes in the skyrmion radius $R$.

The inhomogeneous magnetic structure of the MSH system also allows us to reveal an intriguing connection between the Chern number density, $C({\bf r})$, which is a local marker for the topological nature of the system, and the Berry curvature in momentum space. In particular, the spatial structure of $C({\bf r})$ (see Fig.~\ref{fig:fig1}{\bf d}) reflects that of the skyrmion lattice, but is complementary to that of the induced $\alpha({\bf r})$ (see Fig.~\ref{fig:fig1}{\bf b}), with the maximum in $C({\bf r})$ occurring at the corners of the skyrmion lattice unit cell. However, it is in these regions that the lowest energy states possess their largest spectral weight (see discussion below), establishing a real space analogue of the observation that the lowest energy states in momentum space in general possess the largest Berry curvature.

\begin{figure}[t]
\centering
\includegraphics[width=8.5cm]{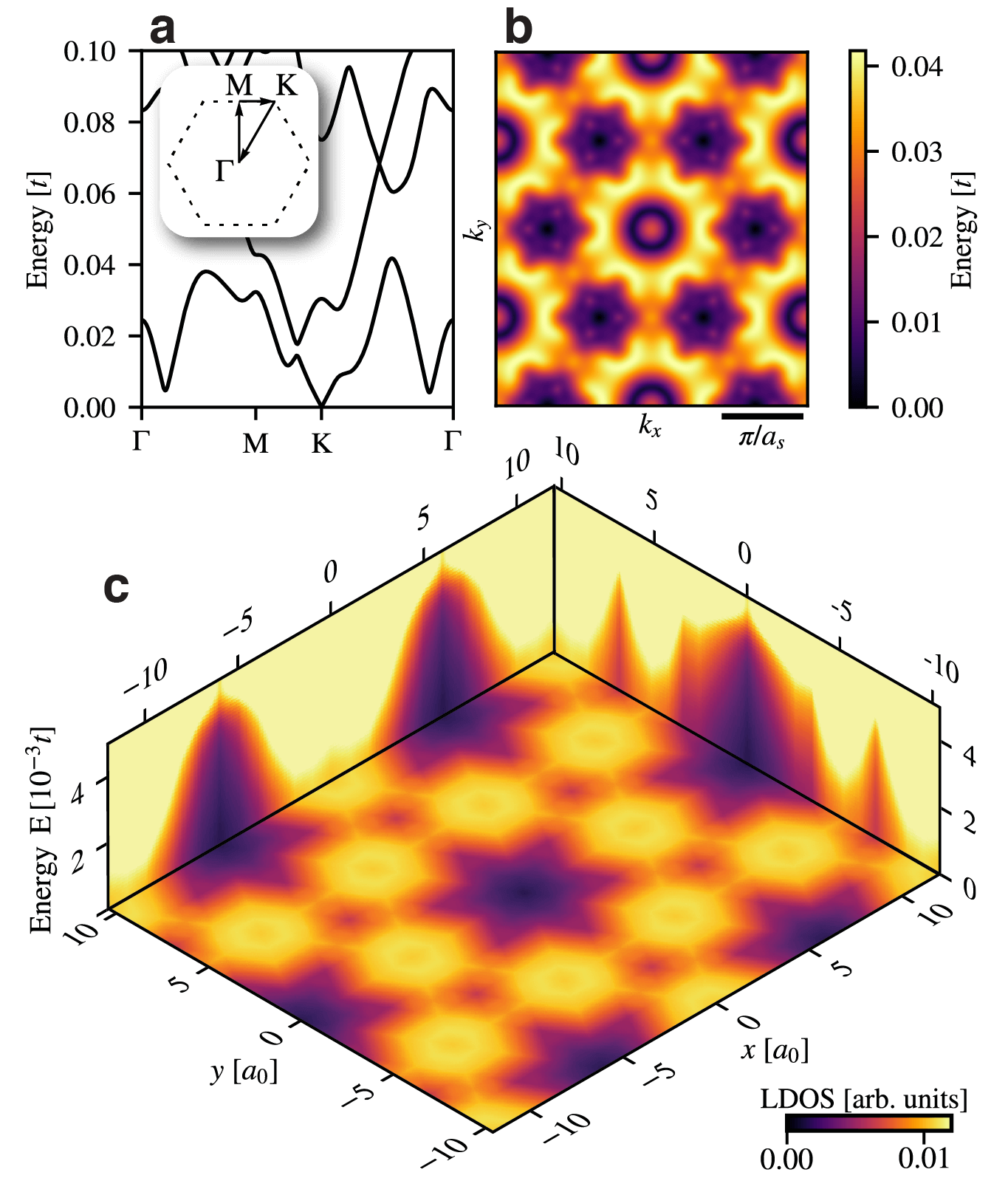}
\caption
{Electronic band structure at a topological phase transition. {\bf a} Electronic bands at the phase transition between two topological phases with $C=6$ and $C=8$ (as indicated by the solid black dot in Fig.~2{\bf c}) for $R=6a_0$ and parameters $(\mu, \Delta, JS)=(-5,0.4,0.657)$t. Shown is the Brillouin zone (BZ) of the skyrmion lattice, i.e., the reduced BZ (RBZ) of the underlying surface lattice. Spatial plot of {\bf b} the dispersion $E_{\bf k}$ of the lowest energy band in the RBZ ($a_s=2R$ is the lattice constant of the skyrmion lattice), and {\bf c} the LDOS at the phase transition, as a function of position and energy.}
\label{fig:fig3}
\end{figure}

\begin{figure*}[t]
\centering
\includegraphics[width=17.cm]{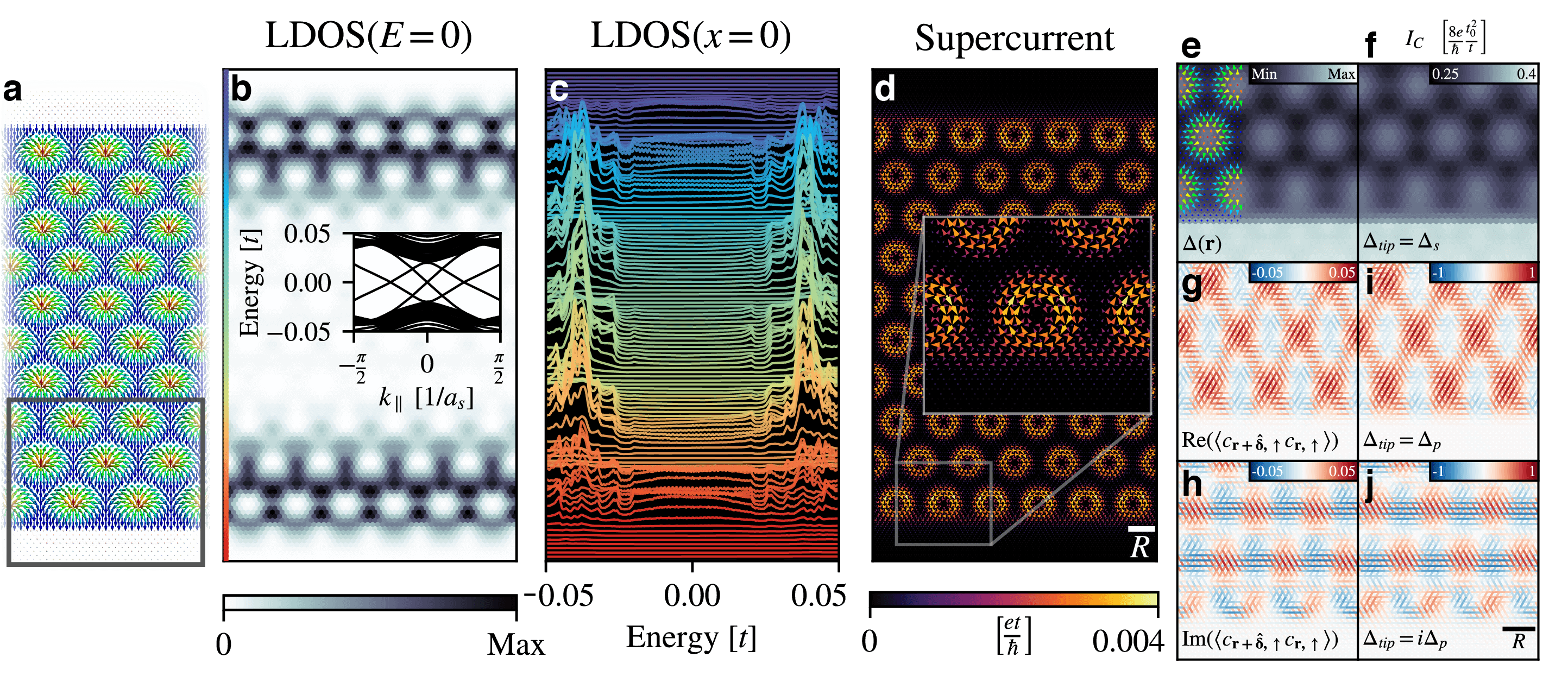}
\caption
{MSH system with a skyrmion ribbon. {\bf a} Schematic form of a skyrmion lattice ribbon on the surface of an $s$-wave superconductors. {\bf b} Spatial plot of the zero energy LDOS. Inset: electronic band structure as a function of momentum, $k_\parallel$ along the ribbon with a width of 27 skyrmions in the $C=3$ phase.  {\bf c} Energy-dependent LDOS for positions from the bottom to the top of the ribbon at $x = 0$ as shown in {\bf b}. {\bf d}  Persistent supercurrents in the skyrmion ribbon. Spatial structure of {\bf e} the superconducting s-wave order parameter, $\Delta({\bf r})$, and {\bf f} the critical Josephson current $I_c$ measured via JSTS using an $s$-wave order parameter in the tip (shown area corresponds to black square in {\bf a}). Spatial structure of the {\bf g} real and {\bf h} imaginary part of the superconducting triplet correlations between nearest neighbor sites. Spatial structure of $I_c({\bf r})$ measured via JSTS with the tip possessing a {\bf i} purely real, and {\bf j} purely imaginary triplet superconduting order parameter. Parameters are $(JS,\Delta,\mu)=(0.5,0.4,-5.5)t$ and $R=5a_0$, corresponding to the $C=3$ phase, with a width of 10 skyrmions.}
\label{fig:fig4}
\end{figure*}

\noindent {\bf Electronic structure at a topological phase transition}
The real space structure of the induced RSO interaction, and hence that of the local topological skyrmion charge, is reflected in the electronic structure of the MSH system, and becomes particularly evident at a topological phase transition.  To demonstrate this, we consider the transition between a $C=8$ and $C=6$ phase, as indicated by the solid black dot in Fig.~\ref{fig:fig2}{\bf c}. While the system possesses a topological gap on either side of the transition, the gap at the transition closes at the $K,K^\prime$-points (see Fig.~\ref{fig:fig3}{\bf a}), as confirmed by a plot of the dispersion $E_{\bf k}$ of the lowest energy band (see Fig.~\ref{fig:fig3}{\bf b}) in the reduced Brillouin zone (RBZ). This gap closing is reflected in a unique spatial and energy structure of the zero-energy local density of states (LDOS) [see $(xy)$-plane in Fig.~\ref{fig:fig3}{\bf c}].
In particular, the spatial structure of the LDOS reveals that the largest (smallest) spectral weight of the zero-energy state, associated with the phase transition, is located where the induced RSO interaction is the smallest (largest), at the corners of the Wigner-Seitz unit cell (the skyrmion center). Thus, the spatial pattern of the zero-energy LDOS is complementary to that of the local topological skyrmion charge, $n_s({\bf r})$. Moreover, as the topological gap in general increases with increasing RSO interaction, we find that the large induced RSO interaction in the skyrmion center leads to a dome-like region in energy in which the LDOS is suppressed [see $(x,E)$- and $(y,E)$-planes in Fig.~\ref{fig:fig3}{\bf c}]. The electronic structure of the skyrmion MSH systems also provides a unique example to demonstrate that the multiplicity $m$ of the momenta in the Brillouin zone, at which the gap closing occurs, determines and is equal to the change in the Chern number at the transition. For
the time-reversal invariant (TRI) $\Gamma, M, (K, K')$ points, the multiplicity is $m=1,3,2$ (note that by symmetry, a gap closing at the $K$ point implies a gap closing at $K^\prime$ as well), respectively, as each $M$ $(K,K^\prime)$ point is shared by 2 (3) BZs, leading to a change in the Chern number by $\Delta C = 1, 3, 2$ at the transition. Gap closings can also occur at non-TRI points, e.g., at points along the $\Gamma - M$ line (see SI section 3), which possess a multiplicity of $m=6$,  resulting in a change of the Chern number by $\Delta C = 6$. While all of the above gap closings exhibit a Dirac cone (see Fig.~\ref{fig:fig3}{\bf a}), there also exist gap closings that exhibit lines of zero-energy (see SI section 3), rather than discrete zero energy Dirac points. These gap closings, however, are not accompanied by a change in the Chern number.

\noindent {\bf MSH system with a skyrmion ribbon} To study the emergence of chiral Majorana edge modes in a skyrmion MSH system, we next consider a skyrmion ribbon placed on the surface of an $s$-wave superconductor (see Fig.~\ref{fig:fig4}{\bf a}). In a topological phase with Chern number $C$, the bulk-boundary requires that such a MSH system possess $|C|$ chiral Majorana edge modes per edge. These modes traverse the superconducting gap and disperse linearly near the chemical potential as a function of the momentum along the ribbon edge, as shown in the inset of Fig.~\ref{fig:fig4}{\bf b} for the $C=3$ phase. A spatial plot of the zero-energy LDOS (see Fig.~\ref{fig:fig4}{\bf b}) demonstrates that the chiral Majorana mode is as expected localized along the edges of the ribbon, and that its spatial structure is complementary to that of the local skyrmion topological charge. The spatial structure of the skyrmion lattice, and hence of the induced $\alpha({\bf r})$, is also reflected in the combined energy and spatial dependence of the LDOS (see Fig.~\ref{fig:fig4}{\bf c}) as revealed by a line-cut of the LDOS from the bottom to the top of the ribbon along $x=0$ (left edge of Fig.~\ref{fig:fig4}{\bf b}). In particular, in the center of the skyrmions, where $\alpha$ is the largest, the spectral weight in the LDOS is pushed to higher energies. The spatial structure of the LDOS is therefore similar to that exhibited by the MSH system at a phase transition (see Fig.~\ref{fig:fig3}{\bf c}).

In addition to the chiral Majorana edge modes, the magnetic structure of the skyrmion ribbon leads to two unique physical features. The first one is the spatial form of persistent supercurrents that are induced by the broken time-reversal symmetry. While these supercurrents are generally confined to the edges of an MSH system \cite{Rac17},
the inhomogeneous magnetic structure of the skyrmion lattice leads to supercurrents (see SI section 4) that circulate each skyrmion, not only along the ribbon's edge, but also in its interior (see Fig.~\ref{fig:fig4}{\bf d}). These supercurrents screen the out-of-plane component of the local magnetic moments, similar to the case of a vortex lattice, and are carried by both the in-gap and bulk states. The second unique feature is the presence of spin-triplet superconducting correlations which are a crucial requirement for the emergence of topological superconductivity \cite{Rac17}. The development of Josephson scanning tunneling spectroscopy (JSTS) \cite{Rod04,Ham15,Ran16,Jae16,Cho19} has provided a unique opportunity to visualize not only these correlations in real space at the atomic level, but also to investigate the effects of the inhomogeneous magnetic structure of the skyrmion lattice on the superconducting $s$-wave order parameter, $\Delta({\bf r})$ \cite{Gra17}. Specifically, pair breaking effects of the magnetic moments lead to a spatially non-uniform suppression of $\Delta({\bf r})$ inside the skyrmion ribbon (see Fig.~\ref{fig:fig4}{\bf e}), with the largest suppression occurring where the induced RSO interaction is the weakest. This spatial structure of $\Delta({\bf r})$ is well imaged by that of the critical Josephson current, $I_c({\bf r})$ (see Fig.~\ref{fig:fig4}{\bf f}), measured via JSTS using a tip with an $s$-wave superconducting order parameter \cite{Gra17}, thus providing direct insight into the strength of local pair breaking effects. Moreover, the inhomogeneous magnetic structure of the skyrmion lattice enables the emergence of superconducting spin-triplet correlations not only in the equal-spin channels $\left\vert \uparrow \uparrow \right\rangle$ and  $\left\vert \downarrow \downarrow \right\rangle$ (corresponding to Cooper pair spin states $S_z = \pm 1$), but also in the mixed-spin $(S_z=0)$ channel, $\left\vert \uparrow \downarrow \right\rangle + \left\vert \downarrow \uparrow \right\rangle$ (see SI section 5). The spatial structure of the real and imaginary parts of these correlations in the $\left\vert \uparrow \uparrow \right\rangle$ channel are shown in Figs.~\ref{fig:fig4}{\bf g} and {\bf h}, respectively (the correlations in the $S_z=0,-1$ channels are shown in SI section 5). These correlations are a direct consequence of the magnetic structure of the skyrmions, and thus vanish outside the ribbon. To image the spatial structure of these non-local triplet correlations, we compute $I_c({\bf r})$ using an extended $(2 \times 1)$ JSTS tip with a superconducting triplet order parameter (see SI section 5). If the tip's order parameter is chosen to be either purely real (see Fig.~\ref{fig:fig4}{\bf i}) or purely imaginary (see Fig.~\ref{fig:fig4}{\bf j}), the resulting Josephson current very well images the spatial structure of the real or imaginary parts, respectively, of the superconducting triplet correlations. We note that these triplet correlations can be imaged despite the fact that the MSH system possesses neither a long-range nor a local triplet superconducting order parameter. Thus JSTS can provide unprecedented insight into the existence of one of the most crucial aspects of topological superconductivity, the existence of spin-triplet correlations.

\noindent {\bf Discussion}
MSH systems containing a magnetic skyrmion layer are suitable candidate systems to explore a rich topological phase diagram. By varying the skyrmion radius, which can be achieved through the application of an external magnetic field, it is possible to tune these systems between different topological phases, and explore not only their unique properties, but also the transitions between them. The origin of this tunability lies in the spatially inhomogeneous Rashba spin-orbit interaction, which is induced by the magnetic skyrmion lattice and which carries a spatial signature in the zero-energy LDOS that can be observed at a topological phase transition. Skyrmion MSH systems also provide a unique opportunity to employ Josephson scanning tunneling spectroscopy to visualize one of the most fundamental aspects underlying the emergence of topological superconductivity, the existence of induced spin-triplet superconducting correlations. This, in turn, yields a new experimental approach to identifying topological superconducting phases, a possibility that needs to be further explored in future studies. These results demonstrate that the tunability of the magnetic structure in MSH systems opens a new venue for the quantum engineering and exploration of topological superconductivity, and the ability to engineer Majorana zero modes and chiral Majorana edge modes. This raises the intriguing question of whether a tuning of topological phases, similar to the one discussed here, could also be achieved using other non-collinear magnetic structures \cite{Nak13,Chen15,Spe20}.\\

\noindent{\bf Acknowledgements}\\
The authors would like to thank H.\ Kim, A. Kubetzka, T. Posske, K. von Bergmann and R.\ Wiesendanger for stimulating discussions. This work was supported by the U. S. Department of Energy, Office of Science, Basic Energy Sciences, under Award No. DE-FG02-05ER46225 (E.M., S.C., and D.K.M.), the {\it Studienstiftung des deutschen Volkes} (J.B.), and through ARC DP200101118 (S.R.).\\


\begin{thebibliography}{10}
\bibitem{nayak-08rmp1083}
C.~Nayak, S.~H. Simon, A.~Stern, M.~Freedman, S.~Das~Sarma, Rev. Mod.
  Phys.\/ {\bf 80}, 1083 (2008).


\bibitem{mourik-12s1003}
V.~Mourik, K. Zuo, S. M. Frolov, S. R. Plissard, E. P. A. M. Bakkers, and L. P. Kouwenhoven, Science {\bf 336}, 1003 (2012).

\bibitem{nadj-perge-14s602}
S.~Nadj-Perge,  I. K. Drozdov, J. Li1, H. Chen, S. Jeon, J. Seo, A. H. MacDonald, B. A. Bernevig, and A. Yazdan, Science {\bf 346}, 602 (2014).

\bibitem{ruby-15prl197204}
M.~Ruby, F. Pientka, Y. Peng, F. von Oppen, B. W. Heinrich, and K. J. Franke, Phys.~Rev.~Lett.~{\bf 115}, 197204 (2015).

\bibitem{pawlak-15npjqi16035}
R.~Pawlak,  M. Kisiel, J. Klinovaja, T. Meier, S. Kawai, T. Glatzel, D. Loss and E. Meyer, npj Quantum Inf.~{\bf 2}, 16035 (2015).

\bibitem{kim-18sa5251}
H.~Kim, A. Palacio-Morales1, T. Posske, L. Rozsa, K. Palotas, L. Szunyogh, M. Thorwart and R. Wiesendanger, Sci. Adv.~{\bf 4}, eaar5251 (2018).

\bibitem{Wang18} D. Wang, L. Kong, P. Fan, H. Chen, S. Zhu, W. Liu, L. Cao, Y. Sun, S. Du, J. Schneeloch, R. Zhong, G. Gu, L. Fu, H. Ding and H.-J. Gao,  Science {\bf 362}, 333 (2018).

\bibitem{Mac19} T. Machida, Y. Sun, S. Pyon, S. Takeda, Y. Kohsaka, T. Hanaguri, T. Sasagawa,
and T. Tamegai,
Nat. Mater. {\bf 18}, 811 (2019).

\bibitem{Men19} G. C. Menard, A. Mesaros, C. Brun, F. Debontridder, D. Roditchev, P. Simon, and T. Cren, Nat. Commun. {\bf 10},  2587 (2019).


\bibitem{menard16-nc2040}
G.~C. Menard,  S. Guissart, C. Brun, R. T. Leriche, M. Trif, F. Debontridder, D. Demaille, D. Roditchev, P. Simon and T. Cren, Nat. Commun.~{\bf 8}, 2040 (2017).


\bibitem{Pal19}  A. Palacio-Morales, E. Mascot, S. Cocklin, H. Kim, S. Rachel, D. K. Morr, and R. Wiesendanger, Sci. Adv.  {\bf 5},  eaav6600  (2019).


\bibitem{Wang20} Z. Wang, J. O. Rodriguez, M. Graham, G. D. Gu, T. Hughes, D.K. Morr, and V. Madhavan,   Science {\bf 367}, 104 (2020).

\bibitem{Ron15} J. R\"{o}ntynen and T. Ojanen,
Phys. Rev. Lett. {\bf 114}, 236803 (2015).

\bibitem{Li16} J. Li, T.  Neupert, Z.J. Wang, A.H. MacDonald, A. Yazdani, and B.A. Bernevig, Nat. Commun. {\bf 7}, 12297 (2016).

\bibitem{Rac17}
S.~Rachel, E.~Mascot, S.~Cocklin, M.~Vojta, D.~K. Morr, Phys. Rev. B
  {\bf 96}, 205131 (2017).

  \bibitem{Rom15} N. Romming, A. Kubetzka, C. Hanneken, K. von Bergmann, and R. Wiesendanger,  Phys. Rev. Lett. {\bf
114}, 177203 (2015).

\bibitem{Yang16}
G.~Yang, P.~Stano, J.~Klinovaja, D.~Loss, Phys. Rev. B {\bf 93}, 224505
  (2016).

\bibitem{Gar19} M. Garnier, A. Mesaros, and P. Simon, Commun. Phys. {\bf 2} , 1 (2019).

\bibitem{Ryu10}  S. Ryu, A. P. Schnyder, A. Furusaki and A. W. W. Ludwig, New J. Phys. {\bf 12}, 065010 (2010).


\bibitem{Mas19} E. Mascot, S. Cocklin, S. Rachel, and D. K. Morr
Phys. Rev. B {\bf 100}, 184510 (2019).

  \bibitem{bianco-11prb241106} R. Bianco and R. Resta, Phys. Rev. B {\bf 84}, 241106(R) (2011).

\bibitem{prodan11jpa113001}
E.~Prodan, J. Phys. A: Math. Theor.~{\bf 44}, 113001 (2011).

\bibitem{prodan17}
E.~Prodan, in {\it A Computational Non-commutative Geometry Program for Disordered
  Topological Insulators\/}, vol.~23 (Springer Briefs in Mathematical Physics,
  2017).

\bibitem{Chen15} W. Chen and A. P. Schnyder, Phys. Rev. B {\bf 92}, 214502 (2015).


\bibitem{Rod04} J.G. Rodrigo, H. Suderow, and S. Vieira,  Eur. Phys. J. B {\bf 40}, 483 (2004).

\bibitem{Ham15} M. H. Hamidian, S. D. Edkins, Sang Hyun Joo, A. Kostin, H. Eisaki, S. Uchida, M. J. Lawler E.-A. Kim,
A. P. Mackenzie, K. Fujita, Jinho Lee  and J. C. Davis, Nature {\bf 532}, 343 (2016).

\bibitem{Ran16} M. T. Randeria, B. E. Feldman, I. K. Drozdov, and A. Yazdani, Phys. Rev. B {\bf 93}, 161115(R) (2016).

\bibitem{Jae16} B. J\"{a}ck, M. Eltschka, M. Assig, M. Etzkorn, C. R. Ast, and K. Kern,
Phys. Rev. B {\bf 93}, 020504(R) (2016).

\bibitem{Cho19}  D. Cho, K.M. Bastiaans, D. Chatzopoulos, G.D. Gu, and M.P. Allan,  Nature {\bf 571}, 541 (2019).

\bibitem{Gra17} M. Graham and D. K. Morr, Phys. Rev. B {\bf 96}, 184501 (2017).

\bibitem{Nak13} S. Nakosai, Y. Tanaka, and N. Nagaosa, Phys. Rev. B {\bf 88}, 180503(R) (2013).

\bibitem{Spe20} J. Spethmann, S. Meyer, K. von Bergmann, R. Wiesendanger, S. Heinze, and A. Kubetzka,  arXiv:2003.02210.









\end{thebibliography}
\end{document}


\title{{\Large Topological Superconductivity in Skyrmion Lattices\\[0.5cm]}
{\large Supplemental Information}}

\author{Eric Mascot$^{1}$, Jasmin Bedow$^{1,2}$, Martin Graham$^{1}$, Stephan Rachel,$^{3}$, and Dirk K. Morr$^{1}$\\
\vspace{0.75cm} {\it \normalsize{$^{1}$ Department of Physics, University of Illinois at Chicago, Chicago, IL 60607, USA\\
$^{2}$ Lehrstuhl f\"{u}r Theoretische Physik I, Technische Universit\"{a}t Dortmund, 44221 Dortmund, Germany\\
$^{3}$ School of Physics, University of Melbourne, Parkville, VIC 3010, Australia\\}}}

\maketitle

\begin{center}
{\bf Section 1: Spin structure of the skyrmion lattice and the induced Rashba spin-orbit interaction}
\end{center}

The local spin at site ${\bf r}$ in the skyrmion lattice is given by
\begin{equation}
	{\bf S}_{\bf r} = S(\sin\theta_{\vec{r}} \cos\phi_{\vec{r}}, \sin\theta_{\vec{r}} \sin\phi_{\vec{r}}, \cos\theta_{\vec{r}})
\end{equation}
with $\theta_{\vec{r}}$ and $\phi_{\vec{r}}$ being the polar and azimuthal angles, defined by
\begin{equation}
	\theta_{\vec{r}} = \pi \min\left(\frac{|{\bf r}-{\bf R_i}|}{R}, 1 \right)
	\quad
	\phi_{\vec{r}} = \arctan\left(\frac{y-y_i}{x-x_i}\right)
\end{equation}
where ${\bf r}=(x,y)$, ${\bf R}_i=(x_i,y_i)$ is the position of the skyrmion center closest to ${\bf r}$, and $R$ is the skyrmion radius.

To demonstrate that the skyrmion lattice induces an effective Rashba spin-orbit interaction, we perform a local unitary transformation that rotates the local spin $\vec{S}_{\vec{r}}$ into the ${\hat z}$-axis, perpendicular to the plane of the MSH system. The corresponding unitary transformation is defined via
\begin{equation}
	\begin{pmatrix}
		c_{\vec{r} \uparrow} \\
		c_{\vec{r} \downarrow}
	\end{pmatrix}
	= {\hat U}_{\vec{r}}
	\begin{pmatrix}
		d_{\vec{r} \uparrow} \\
		d_{\vec{r} \downarrow}
	\end{pmatrix}
\end{equation}
where the unitary matrix ${\hat U}_{\vec{r}} $ is given by
\begin{equation}
	{\hat U}_{\vec{r}} = \exp\left(
		-i \frac{\theta_{\vec{r}}}{2} \frac{\hat{z} \times \vec{S}_{\vec{r}}}{|\hat{z} \times \vec{S}_{\vec{r}}|} \cdot {\bm \sigma}
	\right)
\end{equation}
The Hamiltonian in this basis is then given by
\begin{align}
	H =& \sum_{\vec{r}, \vec{r}', \alpha, \beta} \left(
		-t_{\vec{r}\vec{r}'} {\hat U}_{\vec{r}}^\dagger {\hat U}_{\vec{r}^\prime}
	\right)_{\alpha\beta} d_{\vec{r} \alpha}^\dagger d_{\vec{r} \beta}
	+ \Delta \sum_{\vec{r}} \left(
		d_{\vec{r} \uparrow}^\dagger d_{\vec{r} \downarrow}^\dagger
		+ \mathrm{H.c.}
	\right) + \sum_{\vec{r}, \alpha} (JS \sigma_{\alpha\alpha}^z - \mu) d_{\vec{r} \alpha}^\dagger d_{\vec{r} \alpha}
\end{align}
The effective hopping is
\begin{equation}
	-t_{\vec{r}\vec{r}'} {\hat U}_{\vec{r}}^\dagger {\hat U}_{\vec{r}^\prime}
	= \begin{pmatrix}
		-\tilde{t}_{\vec{r}\vec{r}'} &
		-\alpha_{\vec{r}\vec{r}'}^* \\
		\alpha_{\vec{r}\vec{r}'} &
		-\tilde{t}_{\vec{r}\vec{r}'}^*
	\end{pmatrix} \ ,
\end{equation}
with \(\alpha_{\vec{r}\vec{r}'}\) being the induced Rashba spin-orbit interaction. \\

\begin{center}
{\bf Section 2: Chern number density in real space}
\end{center}

Insight into the interplay between the inhomogeneous magnetic structure of the skyrmion lattice and the local topological nature of the system can be gained by considering the  spatial Chern number density \cite{Bia11,Pro11,Pro17,Mas19,Mas19a} given by
\begin{equation}
C(\textbf{r}) = \frac{N^2}{2\pi i} {\rm Tr}_{\tau,\sigma} [P[\delta_x P, \delta_y P]]_{\textbf{r},\textbf{r}}\ .
\end{equation}
The Chern number is then computed via $C = 1/N^2 \sum_{\textbf{r}}  C(\textbf{r})$, and coincides with that obtained from Eq.(2) in the main text. Here, ${\rm Tr}_{\tau,\sigma}$ denotes the partial trace over spin $\sigma$ and Nambu space $\tau$,
\begin{equation}
\delta_i P= \sum_{m=-Q}^Q c_m e^{-2\pi i m \hat x_i/ N} P e^{2\pi i m \hat x_i / N}\ ,
\end{equation}
where the projector $P$ onto the occupied bands is given by $P=\sum_{\alpha={\rm occ.}} \ket{\psi_\alpha}\bra{\psi_\alpha}$ for a real-space $N\times N$ lattice. The  $c_m$'s are central finite difference coefficients for approximating the partial derivatives. The coefficients for positive $m$ can be calculated by solving the following linear set of equations for $c = (c_1,\ldots,c_Q)$: $\hat A c = b, A_{ij} = 2j^{2i-1}, b_i = \delta_{i,1}, i,j \in \{1,...,Q\}$. For negative $m$, we have $c_{-m} = -c_m$. We have taken the largest possible value of $Q= N/2$.  $C(\textbf{r})$ as defined above thus represents the real-space analog of the Berry curvature $\mathcal{F}({\bf k})$, and was previously introduced to discuss the topological phases of Chern insulators\,\cite{Bia11} and of disordered topological superconductors \cite{Mas19a}.

An alternative form of the Chern number density can be derived by realizing that the periodicity of the skyrmion lattice can be employed to rewrite the Hamiltonian of Eq.(1) of the main text in the following form (we consider a hopping $-t$ between nearest-neighbor sites only, with ${\bm \delta}$ being the lattice vector between nearest neighbor sites)
\begin{align}
	H =& \sum_{\bf{k}}^{\mathrm{RBZ}} \left\{ \left. \sum_{\bf{r}}\right.^{\prime} \left[ \sum_{{\bm \delta}, \sigma} \left( -t e^{i\vec{k} \cdot {\bm \delta}} \right)
	c_{\bf{k}, \bf{r}, \sigma}^\dagger c_{\vec{k}, \vec{r}+{\bm \delta}, \sigma} - \mu \sum_{\sigma} c_{\bf{k}, \bf{r}, \sigma}^\dagger c_{\vec{k}, \vec{r}, \sigma}
	+ \Delta \left( c_{\vec{k}, \vec{r}, \uparrow}^\dagger c_{-\vec{k}, \vec{r}, \downarrow}^\dagger + \mathrm{H.c.} \right) \right. \right. \nonumber \\
	& \left. \left. + J \sum_{\alpha, \beta} \vec{S}_{\vec{r}} \cdot c_{\vec{k}, \vec{r}, \alpha}^\dagger  \vec{\sigma}_{\alpha\beta} c_{\vec{k}, \vec{r}, \beta} \right] \right\} \nonumber \\
    & \equiv \sum_{\bf{k}}^{\mathrm{RBZ}} \Psi_{\bf k}^\dagger {\hat H}({\bf k}) \Psi_{\bf k} \ .
	\label{eq:HRBZ}
\end{align}
Here, the primed sum runs over all locations of the unit cell of the skyrmion lattice, and $\Psi_{\bf k}^\dagger,\Psi_{\bf k}$ are spinors in Nambu space, spin space, and positions ${\bf r}$ in the unit cell. Moreover, RBZ is the reduced Brillouin zone whose wave-vectors are defined via
\begin{align}
{\bf k} = \frac{m_1}{N_1} {\bf b}_1 + \frac{m_2}{N_2} {\bf b}_2 \qquad \text{with}  \qquad m_i = 0, \ldots, M_i -1
\end{align}
where $M_i$ ($N_i$) is the number of skyrmion unit cells (number of lattice sites) along the lattice vector ${\bf a}_i$ defined via
\begin{equation}
{\bf a}_1 = a_0 (1,0) \hspace{1cm} {\bf a}_2 = a_0 (1/2,\sqrt{3}/2) \ ,
\end{equation}
and ${\bf b}_i$ are the reciprocal lattice vectors given by
\begin{equation}
{\bf b}_1 = \frac{2 \pi}{a_0} (1,-1\sqrt{3}) \hspace{1cm} {\bf b}_2 = \frac{2\pi}{a_0} (0,2/\sqrt{3}) \ .
\end{equation}
Finally, the fermionic operators are defined via
\begin{subequations}
\begin{align}
c_{\bf{k}, \bf{r}, \sigma}^\dagger  &= \frac{1}{\sqrt{L_1 L_2}} \sum_{\bf q} c_{\bf{k+q}, \sigma}^\dagger e^{i {\bf q}\bf{r}} \\
c_{\bf{p},\sigma}^\dagger  &= \frac{1}{\sqrt{N}} \sum_{\bf r} c_{\bf{r}, \sigma}^\dagger e^{-i {\bf p}\bf{r}}
\end{align}
\end{subequations}
where
\begin{subequations}
\begin{align}
{\bf q} &= \frac{l_1}{L_1} {\bf b}_1 + \frac{l_2}{L_2} {\bf b}_2 \qquad \text{with}  \qquad l_i = 0, \ldots, L_i -1 \\
{\bf p} &= \frac{n_1}{N_1} {\bf b}_1 + \frac{n_2}{N_2} {\bf b}_2 \qquad \text{with}  \qquad n_i = 0, \ldots, N_i -1 \ ,
\end{align}
\end{subequations}
with $N=N_1N_2$, and $L_i$ is the length of the skyrmion unit cell (in units of $a_0$) along ${\bf a_i}$.

In analogy to the calculation of the Chern number in Eq.(2) of the main text, we can now define the Chern number density (i.e., a ''local" Chern number), by diagonalizing ${\hat H}({\bf k })$ in Eq.(\ref{eq:HRBZ}), yielding
\begin{subequations}
\begin{align}
	C(\vec{r}) &= \frac{L_1 L_2}{2\pi i} \int_{\mathrm{RBZ}} d^2k \, \sum_{\tau, \sigma} \langle {\bf r}, \tau, \sigma | P_{\vec{k}} [ \partial_{k_x} P_{\vec{k}}, \partial_{k_y} P_{\vec{k}} ] | \vec{r}, \tau, \sigma \rangle \\
 P_{\bf{k}} & = \sum_{E_n(\bf{k}) < 0} |\Psi_n({\bf k})  \rangle \langle \Psi_n(\bf{k})| \ ,
\end{align}
\end{subequations}
where $E_n({\bf{k}})$ and $|\Psi_n(\bf{k}) \rangle$ are the eigenenergies and eigenvectors of ${\hat H}({\bf k})$, respectively, with $n$ being a band index, and the summation is taken over Nambu and spin-space.\\

 \begin{center}
{\bf Section 3: Topological Phase Transitions and Gap Closings}
\end{center}

\begin{figure*}[t]
\centering
\includegraphics[width=17.cm]{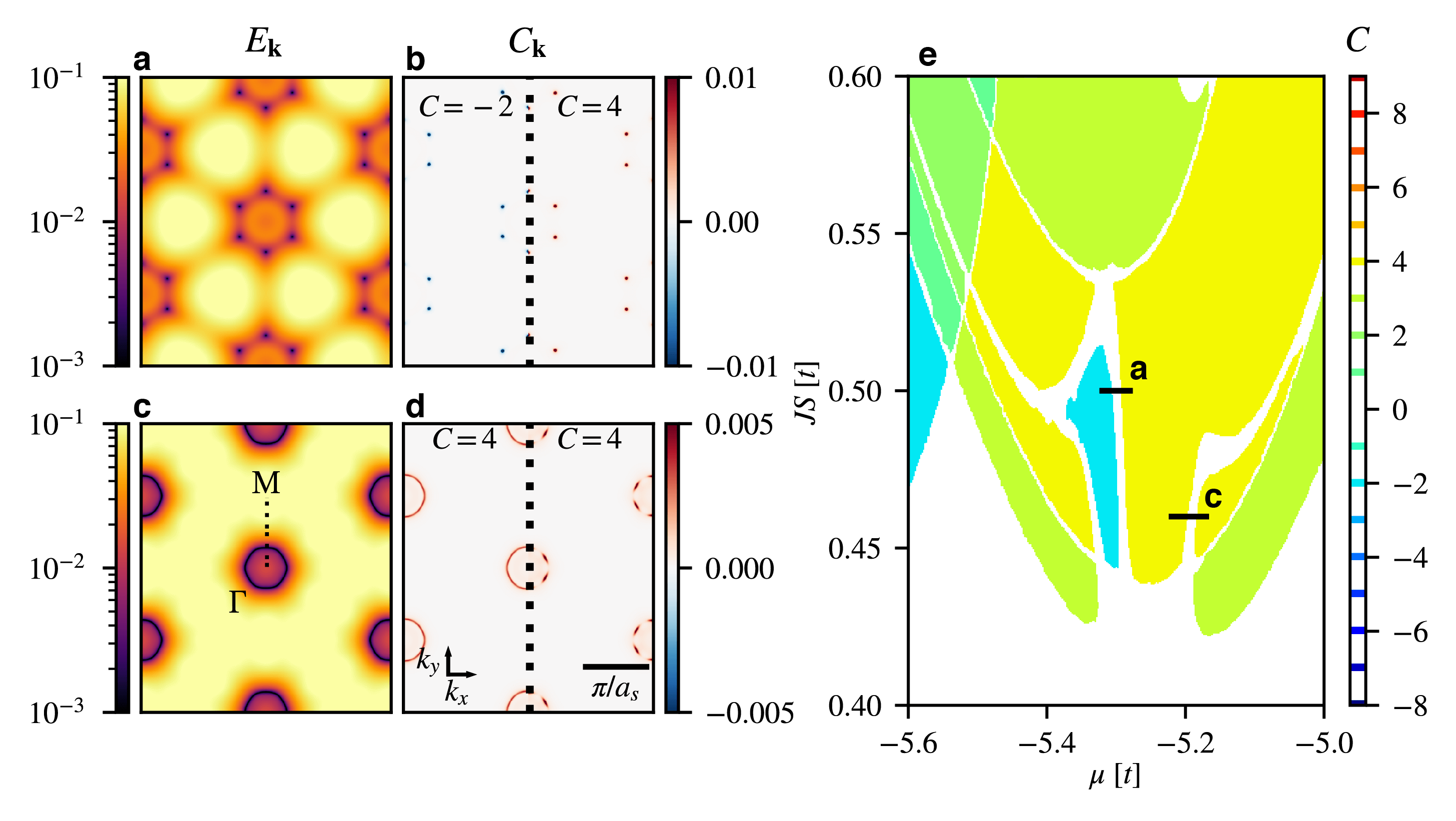}
\caption
{{\bf a} Spatial plot of the dispersion $E_{\bf k}$ of the lowest energy band at the phase transition between the $C=-2$ and $C=4$ phases (see solid black line labelled {\bf a} in {\bf e}). {\bf b} Momentum dependence of $C({\bf k})$ (see text) on opposite sides of the transition in {\bf a}. {\bf c} Spatial plot of the dispersion $E_{\bf k}$ of the lowest energy band for a gap closing without a phase transition (see solid black line labelled {\bf c} in {\bf e}) and {\bf d} the corresponding $C({\bf k})$ on opposite sides of the gap closing. {\bf e} Topological phase diagram with the black lines indicating the gap closings presented in {\bf a} and {\bf c}. Parameters are $R=5 a_0$  and $\Delta = 0.4t$.}
\label{fig:SIfig1}
\end{figure*}

In the topological phase diagram of the skyrmion MSH system, we observe two different types of gap closings. The first one occurs via a Dirac cone, with the Dirac point being located at (a) time-reversal invariant (TRI) $\Gamma, M, K, K'$ points in the Brillouin zone (an example of this is shown in Fig.3 of the main text), or (b) at a non-time-reversal invariant point along the $\Gamma-M$ line. An example of the latter is shown in Fig.~\ref{fig:SIfig1}{\bf a}, where we present the dispersion $E_{\bf k}$ of the lowest energy band at the transition between the $C=4$ and $C=-2$ phases (see black line labelled {\bf a} in Fig.~\ref{fig:SIfig1}{\bf e}). Here, the multiplicity of the points in the BZ where the gap closes is $m=6$ (as none of the points lie on the boundary with another BZ), implying that the Chern number changes by $\Delta C = 6$ at the transition.
To gain further insight into the nature of the topological phase transition, we rewrite the expression for the Chern number in Eq.(2) of the main text as
\begin{align}
 C & =  \frac{1}{2\pi i} \int_{\text{BZ}} d^2k \mathrm{Tr} ( P_{\bf{k}} [ \partial_{k_x} P_{\bf{k}}, \partial_{k_y} P_{\bf{k}} ] ) \equiv \int d^2k \ C({\bf k})
\end{align}
and present the momentum dependence of $C({\bf k})$ in  Fig.~\ref{fig:SIfig1}{\bf b} on either side of the transition. This plot reveals that the change in the Chern number by $\Delta C = 6$ arises from the same change in $C({\bf k})$ near each of the gap closing points. The second type of gap closing in the phase diagram (see black line labelled {\bf c} in  Fig.~\ref{fig:SIfig1}{\bf e}) involves a line of momenta along which the gap vanishes (and thus not discrete Dirac points) in the Brillouin zone, as shown in  Fig.~\ref{fig:SIfig1}{\bf c}. Such a gap closing does not lead to a change in the Chern number between the topological phases adjacent to it, which is also reflected in the momentum dependence of $C({\bf k})$ (see  Fig.~\ref{fig:SIfig1}{\bf d}), which does not show any qualitative change between the two phases.\\[0.1cm]

\begin{center}
{\bf Section 4: Supercurrents in the skyrmion lattice}
\end{center}

The persistent supercurrent associated with the hopping of an electron with spin $\sigma$ from site ${\bf r}_s$ to a nearest neighbor site ${\bf r}_s+{{\bm \delta}}$ can be computed via
\begin{align}
 I^\sigma_{{\bf r}_s,  {\bf r}_s+{ {\bm \delta}}} &= - \frac{2e}{\hbar} (-t) \int \frac{d\omega}{2\pi} \text{Re} [ g^{<}_{mn}({\bf r}_s, {\bf r}_s+{ {\bm \delta}})] \ ,
 \label{eq:I}
\end{align}
where $-t$ is the spin-independent hopping parameter between sites ${\bf r}$ and ${\bf r}_s+{{\bm \delta}}$, and
${g}^{r,a,<}_{mn}({\bf r},{\bf r}+{ {\bm \delta}},\omega)$ are the $(m,n)$ elements in Nambu space of the retarded, advanced, or lesser Green's function matrices. The Green's function matrix in Matsubara time is defined via
\begin{align}
  {\hat g}({\bf r}_s,{\bf r}_s+{{\bm \delta}},\tau) & = -\langle T_\tau \Psi_{{\bf r}_s}(\tau) \Psi_{{\bf r}_s+{ {\bm \delta}}}^\dagger (0) \rangle
  \label{eq:gf}
\end{align}
where the spinors are defined via
\begin{align}
\Psi_{{\bf r}_s}^\dagger &= \left ( c^\dagger_{{\bf r}_s,\uparrow}, c^\dagger_{{\bf r}_s,\downarrow}, c_{{\bf r}_s,\downarrow},c_{{\bf r}_s,\uparrow} \right ) \ .
\label{eq:spinor}
\end{align}
To obtain the above Greens functions for the system, we diagonalize the real space Hamiltonian in Eq.(1) of the main text, yielding energy eigenvalues $E_k$ and eigenvectors $u_{mk}({\bf r})$. The Greens functions can then be computed using
\begin{subequations}
\begin{align}
g^{r}_{mn}(\mathbf{r}_s,\mathbf{r}^\prime_s,\omega) &= \sum_{k}  \frac{u_{mk}(\mathbf{r}_s) u_{nk}^*(\mathbf{r}^\prime_s)}{\omega - E_k + i\delta} \\
g^{a}_{mn} (\mathbf{r}_s,\mathbf{r}^\prime_s,\omega) &= \sum_{k}\frac{u_{mk}(\mathbf{r}_s) u_{nk}^*(\mathbf{r}^\prime_s)}{\omega - E_k - i\delta} \\
g^{<}_{mn} (\mathbf{r}_s,\mathbf{r}^\prime_s,\omega) &= - n_F(\omega)\sum_{k} u_{mk}(\mathbf{r}_s) u_{nk}^*(\mathbf{r}^\prime_s) \left ( \frac{1}{\omega - E_k + i\delta} - \frac{1}{\omega - E_k - i\delta}  \right ) \nonumber \nonumber \\
&=  2 \pi i n_F(\omega)\sum_{k} u_{mk}(\mathbf{r}_s) u_{nk}^*(\mathbf{r}^\prime_s)  \delta({\omega - E_k}) \ .
\label{eq:GFs}
\end{align}
\end{subequations}
Note that the energy eigenvalues come in pairs $\pm E_k$, and that the summation in the above equation runs over all eigenvalues. The elements $(m,n)$ of the Greens function matrix used in Eq.(\ref{eq:I}) is determined by the spin index $\sigma$ with $n=m=1$ for  $\sigma= \uparrow$, and $n=m=2$ for  $\sigma= \downarrow$. The total supercurrent between neighboring sites is then given by
\begin{align}
 I_{{\bf r}_s, {\bf r}_s+{ {\bm \delta}}} & = I^\uparrow_{{\bf r}_s, {\bf r}+{ {\bm \delta}}} + I^\downarrow_{{\bf r}_s, {\bf r}+{{\bm \delta}}}
\end{align}
and is presented in Fig.~4{\bf d} of the main text.\\[0.1cm]

\newpage

\begin{center}
{\bf Section 5: Spatial imaging of spin-triplet superconducting correlations via JSTS}
\end{center}

\begin{figure*}[t]
\centering
\includegraphics[width=17.cm]{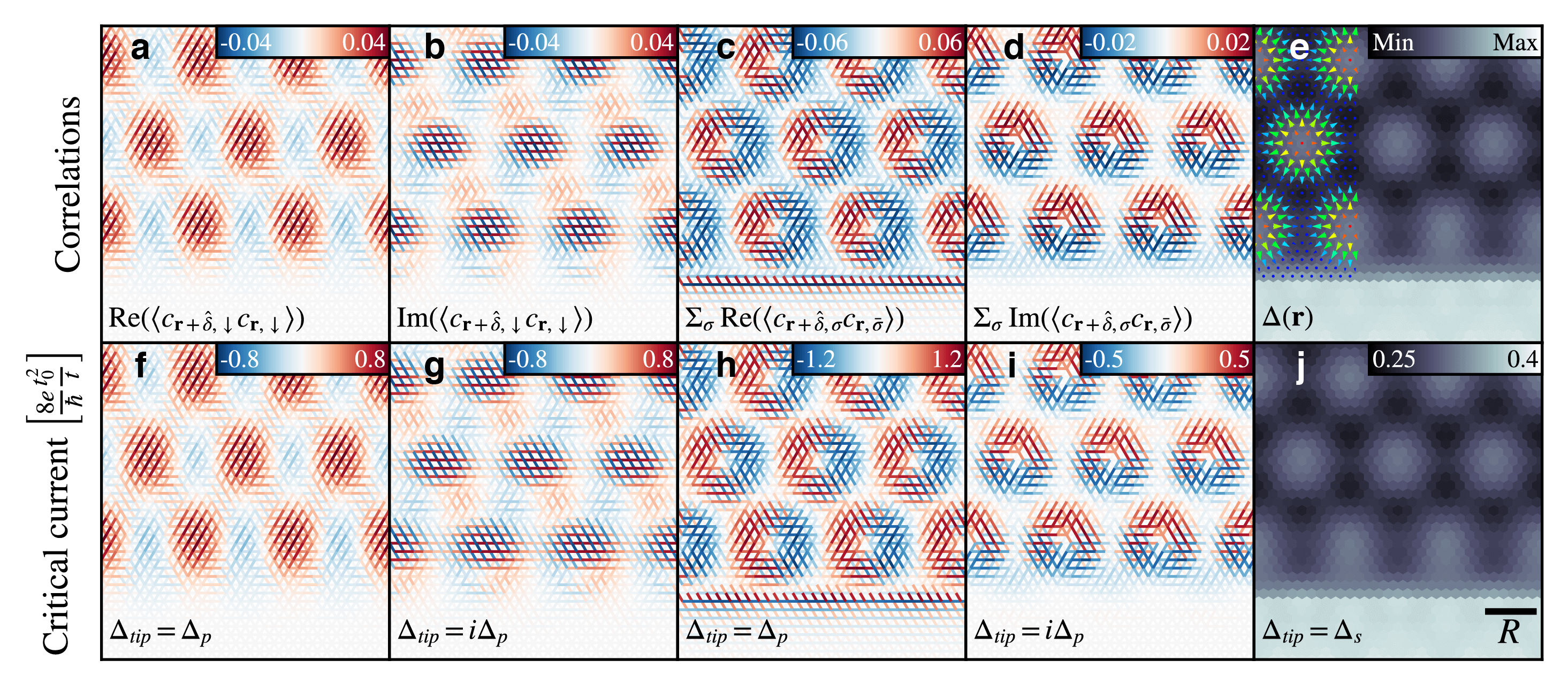}
\caption
{{\bf a}-{\bf d} Spatial structure of the superconducting triplet correlations between nearest-neighbor sites (red corresponding to positive in +x direction) and {\bf f}-{\bf i} the Josephson current measured with a superconducting spin-triplet JSTS tip (for details, see text). {\bf e} Spatial structure of the local superconducting singlet correlations and {\bf j} the Josephson current measured with an s-wave JSTS tip as shown in Fig.~4 of the main text for comparison. Parameters used are $(JS, \Delta, \mu) = (0.5, 0.4, -5)t$ and $R=5a_0$.}
\label{fig:SIfig2}
\end{figure*}
The spatial spin-triplet superconducting correlations between nearest-neighbor sites ${\bf r}$ and ${\bf r}+{{\bm \delta}}$ in the equal-spin-pairing states $\left\vert \uparrow \uparrow \right\rangle$ and $\left\vert \downarrow \downarrow \right\rangle$, and in the opposite spin-pairing state $\left\vert \uparrow \downarrow \right\rangle + \left\vert \downarrow \uparrow \right\rangle$ are given by [using the definition of the Greens function matrix and the spinor given in Eqs.(\ref{eq:gf}) and (\ref{eq:spinor}), respectively]
\begin{subequations}
\begin{align}
\langle c_{{\bf r}+{ {\bm \delta}},\uparrow}c_{{\bf r},\uparrow}\rangle & = \frac{1}{2\pi i} \int d\omega \  g^<_{14}({\bf r},{\bf r}+{{\bm \delta}},\omega)  \ ; \\
\langle c_{{\bf r}+{ {\bm \delta}},\downarrow}c_{{\bf r},\downarrow}\rangle & = \frac{1}{2\pi i} \int d\omega \  g^<_{23}({\bf r},{\bf r}+{{\bm \delta}},\omega) \ ;  \\
\langle c_{{\bf r}+{ {\bm \delta}},\uparrow}c_{{\bf r},\downarrow}\rangle  + \langle c_{{\bf r}+{{\bm \delta}},\downarrow}c_{{\bf r},\uparrow}\rangle
& = \frac{1}{2\pi i} \int d\omega \  \left[ g^<_{24}({\bf r},{\bf r}+{{\bm \delta}},\omega) + g^<_{13}({\bf r},{\bf r}+{ {\bm \delta}},\omega) \right] \ .
\end{align}
\end{subequations}
We are considering superconducting correlations between nearest-neighbor sites only, as would be expected, for example, for a triplet superconductor with a $(p_x +ip_y)$-wave symmetry. Similarly, the local spatial correlations in the superconducting spin-singlet channel with $s$-wave symmetry are obtained via
\begin{align}
\langle c_{{\bf r}+{{\bm \delta}},\uparrow}c_{{\bf r},\downarrow}\rangle  - \langle c_{{\bf r}+{{\bm \delta}},\downarrow}c_{{\bf r},\uparrow}\rangle
& = \frac{1}{2\pi i} \int d\omega \  \left[ g^<_{24}({\bf r},{\bf r}+{ {\bm \delta}},\omega) - g^<_{13}({\bf r},{\bf r}+{{\bm \delta}},\omega) \right] \ .
\end{align}
The spatial form of the local superconducting correlations in the $s$-wave channel are shown in Fig.~4{\bf e}, and for the real and imaginary parts of $\langle c_{{\bf r}+{ {\bm \delta}},\uparrow}c_{{\bf r},\uparrow}\rangle$ in Figs.~4{\bf g} and {\bf h}, respectively, of the main text. The real and imaginary parts of $\langle c_{{\bf r}+{ {\bm \delta}},\downarrow}c_{{\bf r},\downarrow}\rangle$ and of $\langle c_{{\bf r}+{{\bm \delta}},\uparrow}c_{{\bf r},\downarrow}\rangle  + \langle c_{{\bf r}+{ {\bm \delta}},\downarrow}c_{{\bf r},\uparrow}\rangle$ are shown in  Figs.~\ref{fig:SIfig2}{\bf a}, {\bf b} and  Figs.~\ref{fig:SIfig2}{\bf c}, {\bf d}, respectively.

To spatially image these correlations via JSTS, the superconducting order parameter in the JSTS tip needs to possess the same symmetry and spin-structure as the correlations we intend to probe. For the spin-singlet, $s$-wave channel, this has previously been described in Ref.~\cite{Gra17}. In this case, a JSTS tip ending in a single site is sufficient to probe the $s$-wave correlations, and the resulting spatial structure of the critical Josephson current, $I_c$, is shown in Fig.~4{\bf f} of the main text.  In contrast, the triplet superconducting correlations are by definition odd in real space, implying that they are non-local. To probe them via JSTS thus requires a tip that also exhibits a non-local superconducting order parameter (similar to the case of a $d_{x^2-y^2}$-wave order parameter, discussed in Ref.~\cite{Gra19}). This implies that the JSTS tip has to end in at least 2 sites from which electrons can tunnel into the system, a case we will consider in the following (we refer to such a JSTS tip as a 2-site tip). As we have shown before \cite{Gra19}, using tips with a larger number of end sites does not change our qualitative conclusions. Assuming equal-spin-pairing states $\left\vert \uparrow \uparrow \right\rangle$ and $\left\vert \downarrow \downarrow \right\rangle$, and the opposite spin-pairing state $\left\vert \uparrow \downarrow \right\rangle + \left\vert \downarrow \uparrow \right\rangle$ in the tip then leads to the following expressions for the critical Josephson current between the tip and the system, respectively,
 \begin{subequations}
 \begin{align}
I_{\uparrow \uparrow}^{c} & = \frac{e}{\hbar} i \sum_{\mathbf{r}_t,\mathbf{r}^\prime_t  \mathbf{r}_s,\mathbf{r}^\prime_s} t_{\mathbf{r}_t,\mathbf{r}_s} t_{\mathbf{r}^\prime_t,\mathbf{r}^\prime_s}
 \int \frac{d\omega}{2\pi} \left[
	 g^{r}_{14}(\mathbf{r}^\prime_t,\mathbf{r}_t,\omega) g^{<}_{41}(\mathbf{r}_s,\mathbf{r}^\prime_s,\omega) + {g}^{<}_{14}(\mathbf{r}^\prime_t,\mathbf{r}_t,\omega) {g}^{a}_{41}(\mathbf{r}_s,\mathbf{r}^\prime_s,\omega) \right. \nonumber \\
	& \hspace{2cm} \left. + {g}^{r}_{14}(\mathbf{r}^\prime_s,\mathbf{r}_s,\omega) {g}^{<}_{41}(\mathbf{r}_t,\mathbf{r}^\prime_t,\omega) +
	{g}^{<}_{14}(\mathbf{r}^\prime_s,\mathbf{r}_s,\omega) {g}^{a}_{41}(\mathbf{r}_t,\mathbf{r}^\prime_t,\omega) \right]  \ ; \\
I_{\downarrow \downarrow}^{c} & = \frac{e}{\hbar} i \sum_{\mathbf{r}_t,\mathbf{r}^\prime_t  \mathbf{r}_s,\mathbf{r}^\prime_s} t_{\mathbf{r}_t,\mathbf{r}_s} t_{\mathbf{r}^\prime_t,\mathbf{r}^\prime_s} \int \frac{d\omega}{2\pi} \left[ g^{r}_{23}(\mathbf{r}^\prime_t,\mathbf{r}_t,\omega) g^{<}_{32}(\mathbf{r}_s,\mathbf{r}^\prime_s,\omega) + {g}^{<}_{23}(\mathbf{r}^\prime_t,\mathbf{r}_t,\omega) {g}^{a}_{32}(\mathbf{r}_s,\mathbf{r}^\prime_s,\omega) \right. \nonumber \\
	& \hspace{5cm} \left. + {g}^{r}_{23}(\mathbf{r}^\prime_s,\mathbf{r}_s,\omega) {g}^{<}_{32}(\mathbf{r}_t,\mathbf{r}^\prime_t,\omega) + 	{g}^{<}_{23}(\mathbf{r}^\prime_s,\mathbf{r}_s,\omega) {g}^{a}_{32}(\mathbf{r}_t,\mathbf{r}^\prime_t,\omega) \right]  \ ; \\
I_{(\uparrow\downarrow+\downarrow\uparrow)}^{c} & = \frac{e}{\hbar}  i \sum_{\mathbf{r}_t,\mathbf{r}^\prime_t  \mathbf{r}_s,\mathbf{r}^\prime_s} t_{\mathbf{r}_t,\mathbf{r}_s} t_{\mathbf{r}^\prime_t,\mathbf{r}^\prime_s}  \int \frac{d\omega}{2\pi} \left[ g^{r}_{24}(\mathbf{r}^\prime_t,\mathbf{r}_t,\omega) g^{<}_{42}(\mathbf{r}_s,\mathbf{r}^\prime_s,\omega) + {g}^{<}_{24}(\mathbf{r}^\prime_t,\mathbf{r}_t,\omega) {g}^{a}_{42}(\mathbf{r}_s,\mathbf{r}^\prime_s,\omega) \right.  \nonumber \\
	& \hspace{1cm} \left. + {g}^{r}_{24}(\mathbf{r}^\prime_s,\mathbf{r}_s,\omega) {g}^{<}_{42}(\mathbf{r}_t,\mathbf{r}^\prime_t,\omega) + 	{g}^{<}_{24}(\mathbf{R}_s,\mathbf{r}_s,\omega) {g}^{a}_{42}(\mathbf{r}_t,\mathbf{R}_t,\omega) + g^{r}_{13}(\mathbf{r}^\prime_t,\mathbf{r}_t,\omega) g^{<}_{31}(\mathbf{r}_s,\mathbf{r}^\prime_s,\omega) \right. \nonumber \\
& \hspace{1cm} \left.  + {g}^{<}_{13}(\mathbf{r}^\prime_t,\mathbf{r}_t,\omega) {g}^{a}_{31}(\mathbf{r}_s,\mathbf{r}^\prime_s,\omega)  + {g}^{r}_{13}(\mathbf{r}^\prime_s,\mathbf{r}_s,\omega) {g}^{<}_{31}(\mathbf{r}_t,\mathbf{r}^\prime_t,\omega) + {g}^{<}_{13}(\mathbf{r}^\prime_s,\mathbf{r}_s,\omega) {g}^{a}_{31}(\mathbf{r}_t,\mathbf{r}^\prime_t,\omega) \ .
\right]
\label{eq:1}
\end{align}
\end{subequations}
Here, the sums over $\mathbf{r}_s, \mathbf{r}^\prime_s$ ($\mathbf{r}_t, \mathbf{r}^\prime_t$) run over all sites in the system (the tip), with a non-zero tunneling amplitude $t_{\mathbf{r}_t,\mathbf{r}_s}$ ($t_{\mathbf{r}^\prime_t,\mathbf{r}^\prime_s}$) between sites $\mathbf{r}_s$ and $\mathbf{r}_t$ (sites $\mathbf{r}^\prime_t$ and $\mathbf{r}^\prime_s$). Thus, the above equations for the critical Josephson currents are valid for an arbitrary number of sites in the tip. Moreover, we take the tunneling amplitude between sites to be the same for all sites, i.e., $t_{\mathbf{r}_t,\mathbf{r}_s}=t_0$ if electrons can tunnel between site $\mathbf{r}_t$ and $\mathbf{r}_s$. Note that all Greens functions in Eq.(S24) are anomalous, non-local Greens functions. Moreover, we assume that the 2-site  JSTS tip is always perfectly aligned with the bond connecting two nearest-neighbor sites in the triangular lattice. Finally, the Josephson current is given by $I_J = I_c \sin(\Delta\phi) $ where $\Delta \phi$ is the spatially-invariant phase difference between the superconducting order parameters in the tip and the system.

To compute the Greens function of the 2-site JSTS tip, we consider for simplicity the two cases where the triplet superconducting order parameter in the tip is either completely real or completely imaginary. Note that in an infinitely large square lattice with a $(p_x + ip_y)$-wave superconducting order parameter, such a change in the phase of the triplet superconducting order parameter can be achieved by rotating the lattice by $\pi/2$.  To compute the non-local anomalous Greens function for the two sites at the tip end, we use
\begin{subequations}
\begin{align}
g^{r,a}_{nm}(\mathbf{r}_t,\mathbf{r}^\prime_t,\omega) &= \int \frac{d^2 k}{(2\pi)^2}  \frac{\Delta(\mathbf{k}) e^{ik_i}}{(\omega \pm i\delta)^2 - \epsilon(\mathbf{k})^2 - |\Delta(\mathbf{k})|^2} \\
g^{<}_{nm}(\mathbf{r}_t,\mathbf{r}^\prime_t,\omega) &= - n_F(\omega) \left[ g^{r}_{nm}(\mathbf{r}_t,\mathbf{r}^\prime_t,\omega) -g^{a}_{nm}(\mathbf{r}_t,\mathbf{r}^\prime_t,\omega) \right]
\end{align}
\end{subequations}
where
\begin{align}
	\epsilon(\mathbf{k})  &= -2t ( \cos k_x + \cos k_y ) - \mu \ ; \qquad
	\Delta(\mathbf{k}) = -i\frac{\Delta_0 }{2}(\sin k_x + i\sin k_y) \label{eq:tip_OP}
\end{align}
with $\mu = -0.3t$, $\Delta_0$ = 0.3t. The indices $(n,m)$ denote the spin structure of the superconducting gap in the JSTS tip. For $k_i=k_x$, the superconducting order parameter along the bond of the 2-site tip is real, for $k_i=k_y$, it is imaginary. In Figs.4{\bf i} and 4{\bf j} of the main text, we denote these two cases by $\Delta_{tip}=\Delta_p$ and $\Delta_{tip}=i\Delta_p$, respectively. Our results imply that by changing the phase of the superconducting order parameter in the tip, we can probe the real and imaginary parts of the triplet superconducting correlations in the system. These results can be straightforwardly generalized to an arbitrary complex order parameter in the tip.

In  Figs.~\ref{fig:SIfig2}{\bf f} and \ref{fig:SIfig2}{\bf g} we present the critical Josephson current in the $\left\vert \downarrow \downarrow \right\rangle$ channel for $\Delta_{tip}=i\Delta_p$ and $\Delta_{tip}=\Delta_p$, respectively, while in  Figs.~\ref{fig:SIfig2}{\bf h} and \ref{fig:SIfig2}{\bf i}, we show $I_{(\uparrow\downarrow+\downarrow\uparrow)}^{c}$  in the opposite spin-pairing state $\left\vert \uparrow \downarrow \right\rangle + \left\vert \downarrow \uparrow \right\rangle$ for $\Delta_{tip}=i\Delta_p$ and $\Delta_{tip}=\Delta_p$, respectively. For the 2-site tip, we plot the corresponding $I_c({\bf r})$ as a bond quantity. For all of these cases, the spatial form of the critical current very well images that of the superconducting triplet correlations. For comparison, we also present the spatial structure of the superconducting  $s$-wave order parameter, $\Delta({\bf r})$ (see  Fig.~\ref{fig:SIfig2}{\bf e}) and the corresponding $I_c({\bf r})$ (see  Fig.~\ref{fig:SIfig2}{\bf j}). \\[0.2cm]